\documentclass[fleqn,twoside,twocolumn,nofootinbib,showkeys]{revtex4} 
\usepackage[nocpr]{ujp} 

\begin{document}
\title[The Effect of Higher-Order Mesonic Interactions]
{THE EFFECT OF HIGHER-ORDER\\ MESONIC INTERACTIONS ON THE
CHIRAL PHASE \\TRANSITION AND THE CRITICAL TEMPERATURE}%
\author{M. Abu-Shady}
\affiliation{Department of Mathematics, Faculty
of Science, Menoufia University}
\address{Egypt}
\email{abu\_shady\_1999@yahoo.com}
\author{H.M. Mansour\,}%
\affiliation{Department of Physics, Faculty of Science, Cairo University}%
\address{Egypt}%
\udk{539} \pacs{11.10.Wx, 11.30.Rd,\\[-3pt] 12.39 Fe} \razd{\secix}

\autorcol{M.\hspace*{0.7mm}Abu-Shady, H.M.\hspace*{0.7mm}Mansour}

\setcounter{page}{925}%

\begin{abstract}
In the present work, higher-order mesonic interactions are included
in the linear sigma model at a finite temperature. The effective
potential is minimized in the calculations of the sigma and pion
effective masses. The field equations have been solved in the
mean-field approximation, by using the extended iteration method at
a finite temperature. The order of chiral phase transition, the
effective sigma and pion masses, and the effective mesonic potential
are investigated as functions of the temperature. We find that the
chiral phase transition satisfies the Goldstone theorem below the
critical temperature point, when the minimization condition is
satisfied in the chiral limit. The value of the critical temperature
is reduced as compared with that of the original model in agreement with
lattice QCD results. The modified model is compared to with models in
other works.
\end{abstract}
 \maketitle

\section{Introduction}\vspace*{2mm}

The QCD phase transition is of great interest because of its possible
occurrence in the early Universe. The QCD phase transition is also of
interest due to its relevance to the heavy-ion collision experiments at RHIC
and LHC. The lattice QCD calculation [1] indicates that the restoration of
the chiral symmetry occurs at a temperature of the order of 150 MeV, which
is the temperature to be tested in these experiments.

The appropriate framework for studying the chiral phase transitions
is the finite-temperature field theory. Within this framework, the
finite-temperature effective potential is an important and often
used theoretical tool. The use of such techniques goes back to the
1970s when Kirzhnits and Linde [2, 3] first proposed that broken
symmetries at zero temperature could be restored at finite
temperatures. There are two main methods of studying the chiral
phase transition, namely, lattice QCD methods and effective field
theories. Lattice QCD can tell us many things about the phase
transition, but it cannot be used to study the phase transition dynamics
[4]. Hence, there is a need for models that have no such
restriction. One of the effective models to describe baryon
properties is the linear sigma model, which was suggested by
Gell--Mann and Levy [5] for nucleons interacting via the sigma
$(\sigma )$ and pion $(\mathbf{\pi })$ exchanges. The linear sigma
model clarifies how the structure of a nucleon respects the
constraints imposed by chiral symmetry. The spontaneous and explicit
chiral symmetry breakings require the existence of the pion mass.
The model and its extensions provide a good description of the
hadron properties at zero temperature, as explained in Refs. [6--10].
At a finite temperature, the model provides a good description of the
phase transition, by using the Hartree approximation [11--14] within
the Cornwall--Jackiw--Tomboulis (CJT) formalism [15].

In recent years, higher-order multiquark interactions have played
an important role in studying the chiral phase transition and
the critical temperature in chiral quark models. The
Nambu--Jona--Lasinio (NJL) model has been extended to six-quark
interactions. The effect of these interactions on the phase
transition has been investigated in Refs. [16, 17]. Kashiwa {\it et}
al. [18] extended the NJL model to eight-quark interactions and
studied the effect of these interactions on the critical temperature
point and the phase transition. In the same way, Deb and
Bhattacharyya [19] investigated eight-quark interactions in the
Polyakov--Nambu--Jona--Lasinio model, when the chemical potential is
included. Higher-order mesonic interactions are investigated to
provide a description of nucleon properties at a finite temperature in
the linear sigma model [20]. In addition, Abu-Shady and Mansour [21]
considered the effect of the quantization of fields at a finite
temperature in the linear sigma model on the behavior of nucleon
properties.

The aim of this work is to study the effect of higher-order mesonic
interactions on the meson masses, the order of chiral phase
transition, and the critical temperature in the linear sigma model
at a finite \mbox{temperature.}

This paper is organized as follows: the linear sigma model with the
effective mesonic potential at a finite temperature is presented in Section 2.
Next, the numerical calculations and the discussion of results are
given in Section 3. Finally, the summary and the conclusion are presented in
Section 4.

\section{Chiral Quark Sigma\\ Model with the Effective Potential}

\subsection{Chiral quark sigma model\\ with the effective normal mesonic
potential}

The interactions of quarks via $\sigma$- and $\mathbf{\pi
}$-meson can be described at a finite temperature as in Ref. [22]. The
Lagrangian density is
\[L\left( r\right) =i\overline{\Psi }\partial _{\mu }\gamma ^{\mu
}\Psi +\frac{ 1}{2}\left( \partial _{\mu }\sigma \partial ^{\mu
}\sigma +\partial _{\mu } \mathbf{\pi }.\partial ^{\mu }\mathbf{\pi
}\right) \,+\]\vspace*{-7mm}
\begin{equation}
+\,g\overline{\Psi }\left( \sigma +i\gamma _{5}\mathbf{\tau
}.\mathbf{\pi }\right) \Psi -U_{1}^{\rm eff}\left( \sigma
,\mathbf{\pi,}T\right),
\end{equation}%
where
\[U_{1}^{\rm eff}\left( \sigma,\mathbf{\pi,}T\right)
=U_{1}^{T(0)}(\sigma, \mathbf{\pi )+}\frac{7\pi
^{2}T^{4}}{90}\,+\]\vspace*{-7mm}
\[+\left(\!\frac{m_{\sigma }^{2}-m_{\pi }^{2} }{24f_{\pi }^{2}}\!\right)\left( \sigma
^{2}+\mathbf{\pi }^{2}\right) T^{2}\,+\]\vspace*{-7mm}
\begin{equation}
+\left(\!\frac{ m_{\sigma }^{2}-m_{\pi }^{2}}{24f_{\pi
}^{2}}\!\right)T^{2}\left( \!\sigma ^{2}+ \mathbf{\pi
}^{2}-\frac{\nu }{2}^{2}\!\right)\!,
\end{equation}
with
\begin{equation}
U_{1}^{T(0)}\left( \sigma,\mathbf{\pi }\right) =\frac{\lambda
^{2}}{4} \left( \sigma ^{2}+\mathbf{\pi }^{2}-\nu ^{2}\right)
^{2}-m_{\pi }^{2}f_{\pi }\sigma.
\end{equation}
In Eq. (2), $U_{1}^{T(0)}\left( \sigma,\mathbf{\pi }\right) $ is the
usual meson-meson interaction potential at the zero-temperature, where $\Psi
,\sigma,$ and $\mathbf{\pi }$ are the quark, sigma, and pion fields,
respectively. The effective $U_{1}^{\rm eff}\left( \sigma,\mathbf{\pi,}%
T\right) $ is derived in details in Ref. [23] and references therein
at the chiral limit. We extended it to the explicit chiral symmetry as
in Ref. [22]. In the mean-field approximation, the meson fields are
treated as time-independent classical fields. This means that we
replace the powers and the products of meson fields by their
corresponding powers and products of their expectation values.
The meson-meson interactions in Eq. (3) lead to the hidden chiral
$SU(2)\times SU(2)$ symmetry with $\sigma \left( r\right) $ taking
on the vacuum expectation value
\begin{equation}
\langle \sigma \rangle =f_{\pi },
\end{equation}
where $f_{\pi }=93$ MeV is the pion decay constant. The final \ term
in Eq. (3) is included to break the chiral symmetry explicitly. This
leads to a partial conservation of the axial-vector isospin current
(PCAC). The parameters $ \lambda ^{2}$ and $\nu ^{2}$ can be
expressed in terms of$\ f_{\pi }$ and the masses of mesons, as
follows:
\begin{equation}
\lambda ^{2}=\frac{m_{\sigma }^{2}-m_{\pi }^{2}}{2f_{\pi }^{2}},
\end{equation}\vspace*{-5mm}
\begin{equation}
\nu ^{2}=f_{\pi }^{2}-\frac{m_{\pi }^{2}}{\lambda ^{2}}.
\end{equation}

\subsection{The effective mesonic potential\\ with higher-order mesonic
interactions}

 In this subsection, we give the effective potential with higher-order
mesonic interactions in the linear sigma model as in Ref. $\left[
20\right] $. The temperature-dependent effective potential in the
one-loop approximation takes the form
\[U_{2}^{\rm eff}(\sigma,\mathbf{\pi,}T\mathbf{)}=U_{2}^{T(0)}(\sigma
,\mathbf{ \pi )+}\frac{7\pi ^{2}T^{4}}{90}\,+\]\vspace*{-7mm}
\[+\left(\!\frac{m_{\sigma }^{2}-m_{\pi }^{2}}{ 24f_{\pi
}^{2}}\!\right)\left( \sigma ^{2}+\mathbf{\pi }^{2}\right)
T^{2}\,+\]\vspace*{-5mm}
\begin{equation}
+\left(\!\frac{ m_{\sigma }^{2}-m_{\pi }^{2}}{24f_{\pi
}^{2}}\!\right)T^{2}\left( \!\sigma ^{2}+ \mathbf{\pi
}^{2}-\frac{\nu }{2}^{2}\!\right)\!.
\end{equation}
Here, the meson potential $U_{2}^{T(0)}$ at the zero temperature takes the
same form as in [20]:
\begin{equation}
U_{2}^{T(0)}\left( \sigma,\mathbf{\pi }\right) ==\frac{\lambda
^{2}}{4} A(\left( \sigma ^{2}+\mathbf{\pi }^{2})^{2}-B\nu
^{2}\right) ^{2}-m_{\pi }^{2}f_{\pi }\sigma.
\end{equation}
Here, the potential satisfies the chiral symmetry as $m_{\pi
}\rightarrow 0$ and has eight-point interactions corresponding to
the four parameters $\lambda ^{2},\nu ^{2},A,$ and $B.$ Applying the
minimizing conditions $\Big\{\frac{
\partial U_{2}^{T(0)}}{\partial \sigma }\Big|_{\sigma =f_{\pi },\mathbf{
\pi }=0}=0$, $\frac{\partial ^{2}U_{2}^{T(0)}}{\partial \sigma
^{2}}\Big|_{\sigma =f_{\pi },\mathbf{\pi }=0}=m_{\sigma }^{2}$,
$\frac{\partial ^{2}U_{2}^{T(0)}}{\partial \mathbf{\pi }^{2}}\Big|
_{\sigma =f_{\pi }, \mathbf{\pi }=0}=m_{\pi }^{2}\Big\}$, and PCAC,
we get
\begin{equation}
A=\frac{m_{\sigma }^{2}-3m_{\pi }^{2}}{4f_{\pi }^{4}\left( m_{\sigma
}^{2}-m_{\pi }^{2}\right) },
\end{equation}
and\vspace*{-3mm}
\begin{equation}
\ B=f_{\pi }^{2}\left( \!1-\frac{2m_{\pi }^{2}\left( m_{\sigma
}^{2}+m_{\pi }^{2}\right) }{\left( m_{\sigma }^{2}-3m_{\pi
}^{2}\right) ^{2}}\!\right) \!.
\end{equation}
By introducing the dimensionless quantities
$x^{2}=\frac{\mathbf{\pi }^{2}}{ f_{\pi }^{2}}$ and
$y^{2}=\frac{\sigma ^{2}}{f_{\pi }^{2}},$ the effective potential
can be \mbox{rewritten as}
\[U_{2}^{\rm eff}(y,x,T\mathbf{)}
=U_{2}^{T(0)}(y,x\mathbf{)+}\frac{7\pi
^{2}T^{4}}{90}\,+\]\vspace*{-7mm}
\[+\left(\!\frac{m_{\sigma }^{2}-m_{\pi }^{2}}{24f_{\pi }^{2}}\!\right )T^{2}\left(
(yf_{\pi })^{2}+(xf_{\pi })^{2}\right) +\]\vspace*{-5mm}
\begin{equation}
+ \left(\!\frac{m_{\sigma }^{2}-m_{\pi }^{2}}{24f_{\pi
}^{2}}\!\right)T^{2}\left(\! (yf_{\pi })^{2}+(xf_{\pi
})^{2}-\frac{\nu }{2}^{2}\!\right) \!.
\end{equation}
The finite-temperature vacuum can be defined by minimizing the
effective potential as
\begin{equation}
\frac{\partial U_{2}^{\rm eff}(\sigma,\mathbf{\pi
,}T\mathbf{)}}{\partial \sigma }\bigg|_{\sigma =\sigma
_{0}(T),\mathbf{\pi }=0}=0.
\end{equation}%
Equation (12) represents the condition necessary to satisfy the spontaneous
breaking of chiral symmetry, thereby satisfying the Goldstone theorem, where
$\sigma _{0}(T)=f_{\pi }(1+\delta (T)).$ We can rewrite Eq. (12) in
dimensionless variables as follows:
\[\frac{\partial U_{2}^{\rm eff}(\sigma,\mathbf{\pi
,}T\mathbf{)}}{\partial \sigma }\bigg|_{\sigma =\sigma
_{0}(T),\mathbf{\pi }=0}=\]\vspace*{-5mm}
\begin{equation}
=\frac{1}{f_{\pi }} \frac{\partial U_{2}^{\rm
eff}(y,x,T\mathbf{)}}{\partial y}\bigg|_{y=y_{0},x=0}=0.
\end{equation}
Using Eq. (11) and Eq. (13), we obtain
\begin{equation}
2A~\lambda ^{2}f_{\pi }^{6}y_{0}^{7}-2AB\lambda ^{2}\nu
^{2}y_{0}^{3}+\frac{1 }{3}\lambda ^{2}T^{2}y_{0}+m_{\pi }^{2}=0.
\end{equation}
Here, $y_{0}=1+\delta (T).$ In the chiral limit ($m_{\pi }=0),$ Eq. (14)
takes the form
\begin{equation}
\frac{1}{4}y_{0}^{6}-\frac{1}{4}y_{0}^{2}+\frac{T^{2}}{6f_{\pi }^{2}}=0.
\end{equation}
The square of the sigma mass is obtained as the second derivative of
the effective potential as in Refs. [23, 24]:\vspace*{-3mm}
\[m_{\sigma }^{2}(T) =\frac{\partial ^{2}U_{2}^{\rm eff}(\sigma
,\mathbf{\pi )}}{
\partial \sigma ^{2}}\bigg|_{\sigma =\sigma _{0}(T),\mathbf{\pi }=0}=\]\vspace*{-7mm}
\[=\,\frac{ 1}{f_{\pi }^{2}}\frac{\partial ^{2}U_{2}^{\rm
eff}(y,x\mathbf{)}}{\partial y^{2}} \bigg |
_{y=y_{0},x=0},\]\vspace*{-7mm}
\begin{equation}
 m_{\sigma }^{2}(T) =14A~\lambda ^{2}f_{\pi
}^{6}y_{0}^{6}-6AB\lambda ^{2}\nu ^{2}f_{\pi
}^{2}y_{0}^{2}+\frac{1}{3}\lambda ^{2}T^{2}.
\end{equation}
In the chiral limit,\vspace*{-2mm}
\begin{equation}
m_{\sigma }^{2}(T)=m_{\sigma }^{2}\left(\!
\frac{7}{4}y_{0}^{6}-\frac{3}{4}
y_{0}^{2}+\frac{1}{6}\frac{T^{2}}{f_{\pi }^{2}}\!\right)\!.
\end{equation}
Similarly,\vspace*{-3mm}
\[m_{\pi }^{2}(T) =\frac{\partial ^{2}U_{2}^{\rm eff}(\sigma,\mathbf{\pi
,}T \mathbf{)}}{\partial \pi ^{2}}\bigg|_{\sigma =\sigma
_{0}(T),\mathbf{\pi } =0}=\]\vspace*{-7mm}
\begin{equation}
=\frac{1}{f_{\pi }^{2}}\frac{\partial ^{2}U_{2}^{\rm
eff}(y,x,T\mathbf{)}}{
\partial x^{2}}\bigg| _{y=y_{0},x=0},
\end{equation}\vspace*{-7mm}
\begin{equation}
m_{\pi }^{2}(T) =2A~\lambda ^{2}f_{\pi }^{6}y_{0}^{6}-2AB\lambda
^{2}\nu ^{2}f_{\pi }^{2}y_{0}^{2}+\frac{1}{3}\lambda ^{2}T^{2}\!.
\end{equation}
In the chiral limit,\vspace*{-2mm}
\begin{equation}
m_{\pi }^{2}(T)=m_{\sigma }^{2}\left(\!
\frac{1}{4}y_{0}^{6}-\frac{1}{4} y_{0}^{2}+\frac{T^{2}}{6f_{\pi
}^{2}}\!\right) \!,
\end{equation}%
Substituting Eq. (15) into Eqs.(17) and (20), we obtain the
following sigma and pion masses in the chiral limit:\vspace*{-3mm}
\begin{equation}
m_{\sigma }^{2}(T)=m_{\sigma }^{2}\left( \!\frac{3}{2}\left(
1+\delta (T)\right) ^{6}-\frac{1}{2}\left( 1+\delta (T)\right)
^{2}\!\right)\!,
\end{equation}\vspace*{-5mm}

\noindent and\vspace*{-2mm}
\begin{equation}
m_{\pi }^{2}(T)=0.
\end{equation}

\noindent In view of Eqs. (21) and (22), we note that the sigma mass is
finite, and the pion mass is zero. Therefore, the Goldstone
theorem is satisfied at low temperatures. This method is used in Ref.
[23]. By substituting $ y_{0}=1+\delta (T)$ into Eq. (14), we obtain
\begin{equation}
\delta (T)=\frac{-2A\lambda ^{2}f_{\pi }^{6}\!+\!2AB\lambda ^{2}\nu
^{2}f_{\pi }^{2}\!-\!\frac{1}{3}\lambda ^{2}T^{2}\!+\!m_{\pi
}^{2}}{14A~\lambda ^{2}f_{\pi }^{6}-6AB\lambda ^{2}\nu ^{2}f_{\pi
}^{2}+\frac{1}{3}\lambda ^{2}T^{2}}.\!\!\!\!
\end{equation}

\section{Discussion of the Results}

\begin{figure}[h!]%
\vskip1mm
\includegraphics[width=8cm]{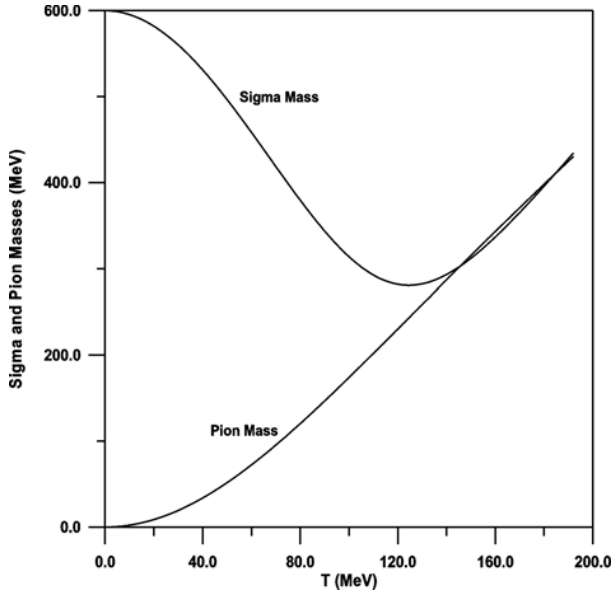}
\vskip-3mm\caption{ Sigma $m_{\sigma }(T)$ and pion $m_{\pi }(T)$
masses as functions of the temperature $T$ for two models in the
chiral limit. The Goldstone theorem is not satisfied }\vskip3mm
\end{figure}

\begin{figure}[h!]%
\includegraphics[width=8cm]{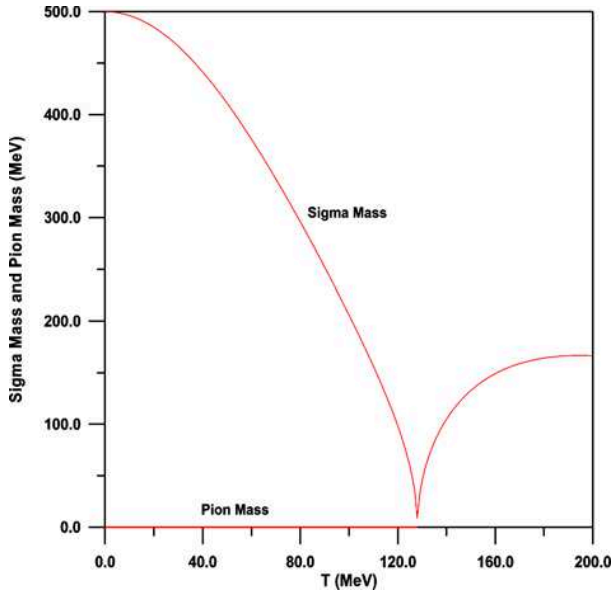}
\vskip-3mm\caption{  Sigma $m_{\sigma }(T)$ and pion $m_{\pi }(T)$
masses are plotted as functions of the temperature $T$ for the
higher-order mesonic model in the chiral limit $(m_{\pi }=0).$ The
Goldstone theorem is satisfied }
\end{figure}

In this work, we concentrate only on the thermal effects and ignore
quantum corrections. It is important to examine the linear sigma
model in the presence of higher-order mesonic interactions. The
higher-order mesonic potential was suggested in Ref. [20] as a good
means of describing the nucleon properties at non-zero temperatures.
So we need to examine the effect of higher-order mesonic
interactions on the meson properties, phase transition, and critical
temperature at non-zero temperatures. We obtained the gap equations
(Eqs.~(16) and (19)) by minimizing the effective mesonic potential
$U_{2}^{\rm eff}\left( \sigma,\mathbf{\pi,T}\right) $. This method
is used in the previous works such as Refs. [23, 24]. For numerical
computation, we used the model parameters at the zero temperature as
the initial conditions; namely, we took $m_{\pi }=140$ MeV,
$m_{\sigma }=500\to$ $\to 700$~MeV, and $f_{\pi }=93$ MeV.

We investigate the effect of the higher-order mesonic contributions
on the sigma and pion masses, the order of phase transition, and the
effective potential. In addition, the effect of the sigma mass as a
free parameter is investigated. In Figs.~1 and 2, we show the
calculated sigma and pion masses in the chiral limit ($m_{\pi }=0)$.
In Fig.~1, the sigma and pion masses are plotted as functions of the
temperature when the minimization condition in Eq. (15) has not been
applied to the sigma and pion masses (Eqs. (16) and (19)). We note
that the sigma mass decreases with increasing temperature, and the
pion mass increases with the temperature. The two masses cross at
the critical temperature, where the restoration of chiral symmetry
is appeared. We note that the Goldstone theorem is not satisfied in
this case. This behavior is in agreement with the findings of other
authors [11, 14], who found that the pion is massive below the
critical temperature, when the Hartree approximation is applied. In
Fig.~2, the effective sigma and pion masses are plotted as functions
of the temperature in the chiral limit. The sigma and pion masses
satisfy the minimization condition in Eq. (15), which leads to a
zero pion mass below the critical temperature and to a sigma mass,
which decreases with increasing temperature. Therefore, the
Goldstone theorem is satisfied. We conclude that the minimization
condition of the spontaneous breaking symmetry is necessary to
satisfy the Goldstone theorem at lower temperatures. This conclusion
agrees with the findings of Nemoto {\it et al.} [24], who showed
that the Goldstone theorem is satisfied when the condition in Eq.
(15) is applied, where the pion mass is defined as the curvature of
an effective potential. Hong {\it et al.} [23] showed also that the
Goldstone theorem is satisfied when the spontaneous breaking of a
symmetry of the system is \mbox{conserved.}\looseness=1

\begin{figure}%
\vskip1mm
\includegraphics[width=\column]{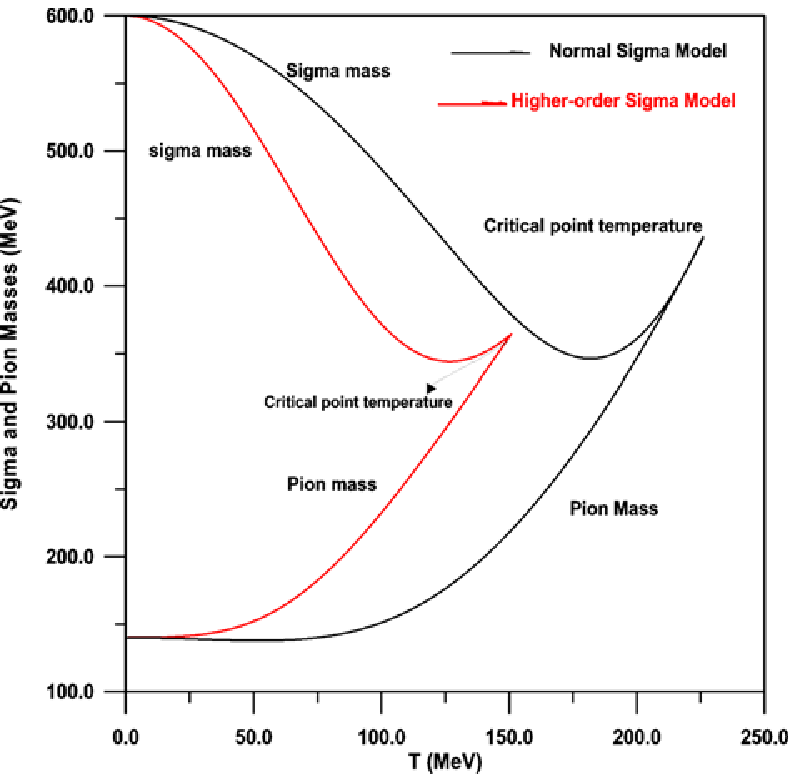}
\vskip-3mm\caption{  Sigma $m_{\sigma }(T)$ and pion $m_{\pi }(T)$
masses are plotted as functions of the temperature $T$ in the above
models in the presence of the explicit symmetry-breaking term. At
temperature $T=$ $=0$, the pions appear with the observed masses
\mbox{($m_{\pi }=140$ MeV)} }
\end{figure}

In Fig. 3, the sigma and pion masses are plotted as functions of the
temperature in the presence of the explicit symmetry-breaking term
$m_{\pi }\neq 0.$ At the zero temperature, the sigma and pion masses
appear with their experimental masses $m_{\sigma }=600$~MeV and
$m_{\pi }=140$~MeV. We note that the inclusion of higher-order
mesonic interactions leads to a reduction in the critical
temperature from 226~MeV to 150~MeV. In Figs.~4 and 5, we
investigate the effect of the sigma mass on the critical point
temperature. In Fig.~4, we note that the critical point temperature
is independent of the sigma mass in the chiral limit. The opposite
is observed in Fig.~5, where we note that the increase in the sigma
mass reduces the value of critical point temperature. Increasing
the sigma mass from $ m_{\sigma }=500$ to 700~MeV corresponds to
decreasing the critical temperature from $ T_{c}=161$~MeV to
$T_{c}=$~127~MeV. The present critical value $ T_{c}=151$ at
$m_{\sigma }=600$~MeV. In comparison, the Wuppertal--Budapest group
$\left[ 26\right] $ found that the transition temperature for the
chiral restoration of $u,d$ quarks equals $T_{c}=151$~MeV.
Therefore, the present result of $T_{c}$ is a good agreement with
lattice QCD results. In addition, we found that %
\begin{figure}[h!]%
\vskip1mm
\includegraphics[width=\column]{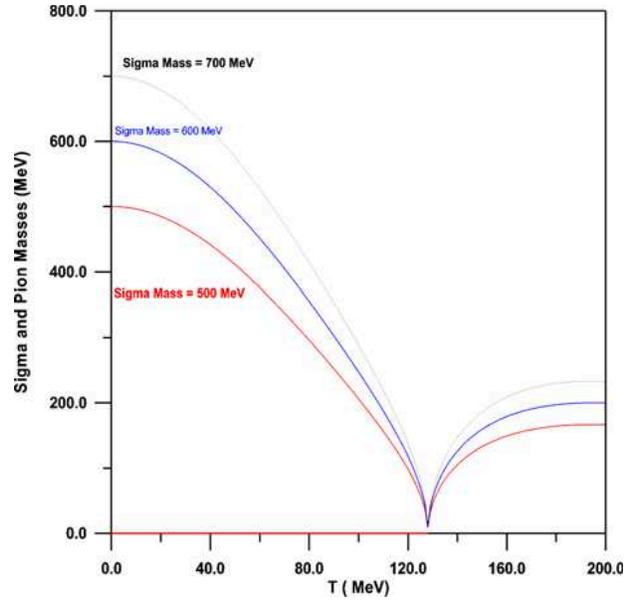}
\vskip-3mm\caption{ Sigma $m_{\sigma }(T)$ and pion $m_{\pi }(T)$
masses are ploted as functions of the temperature in the chiral
limit using the higher-order sigma model for different values of
sigma masses. At the temperature $T = 0$, the pions appear massless
\mbox{($m_{\pi }=0$)}. The Goldstone theorem is satisfied }\vskip9mm
\end{figure}%
\begin{figure}[h!]%
\includegraphics[width=\column]{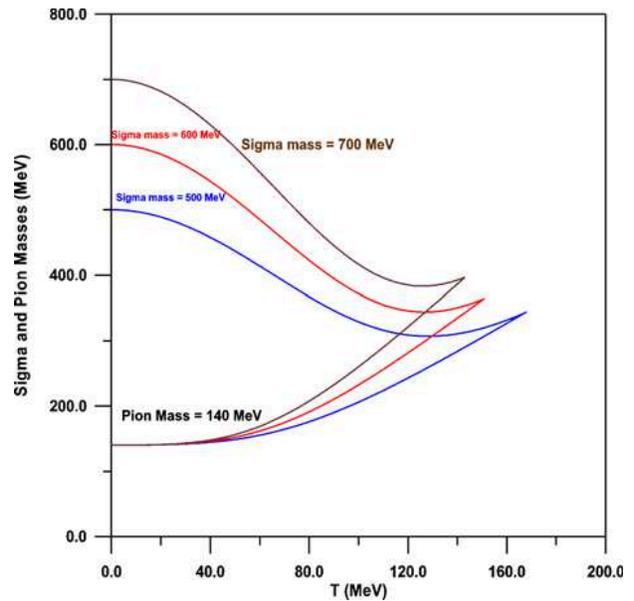}
\vskip-3mm\caption{  Sigma $m_{\sigma }(T)$ and pion $m_{\pi }(T)$
masses are plotted as functions of the temperature in the presence
of the explicit symmetry-breaking term. At the temperature $T = 0$,
the pions appear with the observed masses ($m_{\pi }=140$ MeV)
}\vspace*{-6mm}
\end{figure}%
the Nambu--Jona--Lasinio [18] model, when extended to include
higher-orders of the sigma field ($\sigma ^{4}),$ leads to a
reduction in the critical point temperature to $ T_{c}= 180$~MeV.
Therefore, the  present result is in agreement with the
Nambu--Jona--Lasinio \mbox{model.}\looseness=1

\begin{figure}%
\vskip1mm
\includegraphics[width=7.9cm]{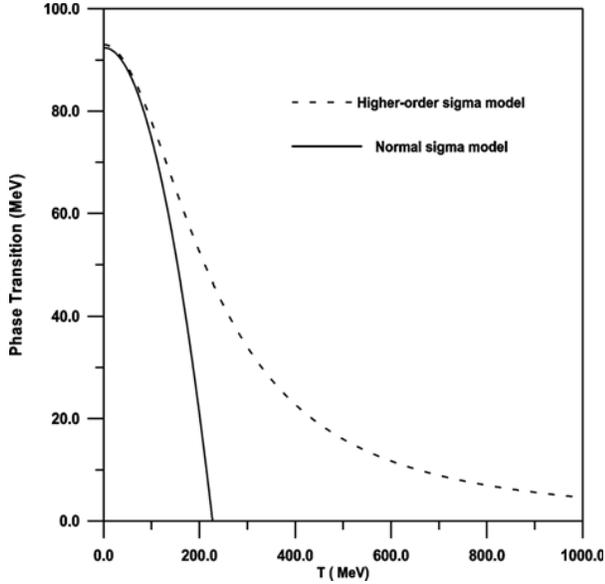}
\vskip-3mm\caption{ The phase transition $\Phi =\left( \sigma
^{2}+\mathbf{\pi } ^{2}\right) ^{1/2}$ is plotted as a function of
the temperature in the normal and higher-order sigma models in the
chiral limit }\vskip1mm
\end{figure}

\begin{figure}%
\includegraphics[width=7.9cm]{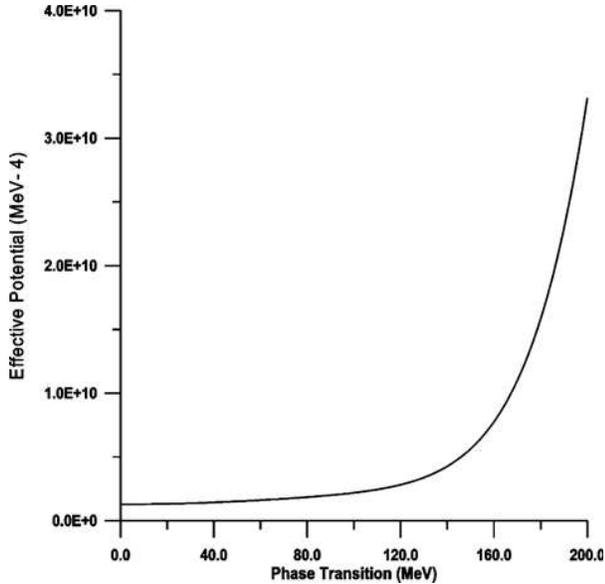}
\vskip-3mm\caption{ The effective mesonic potential $ U_{2}^{\rm
eff}(\sigma,\pi,T)$ is plotted as a function of the phase transition
at the fixed value of temperature $T= 200$ MeV }\vspace{-2mm}
\end{figure}

In Fig.\,\,6, we examine the effect of a finite temperature on the
chiral phase transition $\Phi =$\linebreak $=( \sigma
^{2}+\mathbf{\pi }^{2}) ^{1/2}.$ It is clear that the quantity $\Phi
$ plays the role of the order parameter of the phase transition. In
a previous work [22], we investigated the order of the phase
transition and found that it is equal to 2. We note that the
inclusion of the higher-order mesonic interactions in the normal
sigma model converts the phase transition from a second-order phase
transition to a crossover. The interpretation of the transition
becoming a crossover is given in Ref. [11], which states that the
phase transition vanishes at very high values of temperature.
Kashiwa {\it et al}. [18] showed that the inclusion of a vector
interaction $w^{2}$ makes the phase transition smoother, but the
higher-order interaction $\sigma ^{4}$ makes the phase transition
sharper in the NJL model. In addition, they found that the $w^{2}$
interaction tends to change the phase transition from a first-order
one to a crossover. Moreover, the phase transition is converted from
a first-order transition to a crossover by including the term
$\sigma ^{4}+w^{2}$ in the NJL model. The Wuppertal--Budapest group
[26]  found that the phase transition is a crossover. Therefore, the
result in Fig.~6 gives similar results that the phase transition
changes from a second-order phase transition to a crossover. In
Fig.~7, the higher-order mesonic effective potential is plotted as a
function of the phase transition. We note that the effective mesonic
potential increases with the phase transition. The potential has no
minima at $\Phi \neq 0,$ which can be interpreted as a crossover
[27]. Therefore, we can interpret the phase transition as a
crossover in agreement with the result obtained \mbox{in
Fig.~6.}\looseness=1

In Refs. [11, 14, 24], the authors ignored the effect of the fermion
sector in the linear sigma model at a finite temperature. In this
work, we consider this sector and its effect on the effective
potential. So the effective potential is plotted in the normal
potential and the higher-order potential which are defined by
Eq.~(2) and Eq.~(7), respectively. By comparison of the two contours
in Figs.~8 and 9, we note that the values of effective potential are
reduced in comparison with the normal potential by about 53\% at the
higher values of temperature.  This indicates that the higher-order
mesonic contributions and the quark mass play an important role at
higher values of temperature. This means that increasing the mesonic
contributions in the normal sigma model leads to the increase of the
coupling between the meson fields. In addition, the increase of the
quark mass leads to an increase of the coupling constant between
quark and meson fields. Therefore, this leads to the strong decrease
in the higher-order potential which has an effect on the behavior of
the nucleon mass at finite temperatures and then the deconfinement
phase \mbox{transition [20].}\looseness=1

\section{Comparison with Other Works}

It is interesting to compare the present results with the results of
other groups. Here, we compare our results with the results of the
Nambu--Jona--Lasinio model with its extension and with QCD lattice
calculations [26]. The chiral phase transition was studied with the
use of the NJL model with a Polyakov loop in Refs. [28, 29]. Hansen
{\it et al}. [28] studied the properties of the scalar and
pseudoscalar mesons at finite temperatures and the chemical
potential, by using the PNJL model, and the role of pions as
Goldstone bosons. This means that the Goldstone theorem is satisfied
in their calculations. In the present work, we find that the
Goldstone theorem is satisfied in the chiral limit. So the present
results are in good agreement with Ref. [28]. Deb {\it et al}. [29]
investigated the phase diagram and the location of the critical end
point (CEP), where the CEP is the point that separates the crossover
from the first-order phase transition within the NJL and PNJL
models. They found that the CEP point is shifted to a lower chemical
potential and a higher temperature in the presence of eight-quark
interactions. In the present work, we do not include the chemical
potential, hence the CEP point is not calculated in our model.
Osipov {\it et al}. [17] investigated eight-quark interactions in
the framework of the NJL model and the Hooft Lagrangian. They found
that the critical point temperature is reduced to a lower value when
eight-quark interactions are included. Therefore, the present result
is in agreement with their result. We find that the critical point
temperature is shifted to a lower value when eight-mesonic
interactions are included. Also, Hiller {\it et al}. [30] found that
the critical temperature is shifted to a lower value when six-quark
interactions are considered in the chiral phase transition. %
\begin{figure}%
\vskip1mm
\includegraphics[width=7cm]{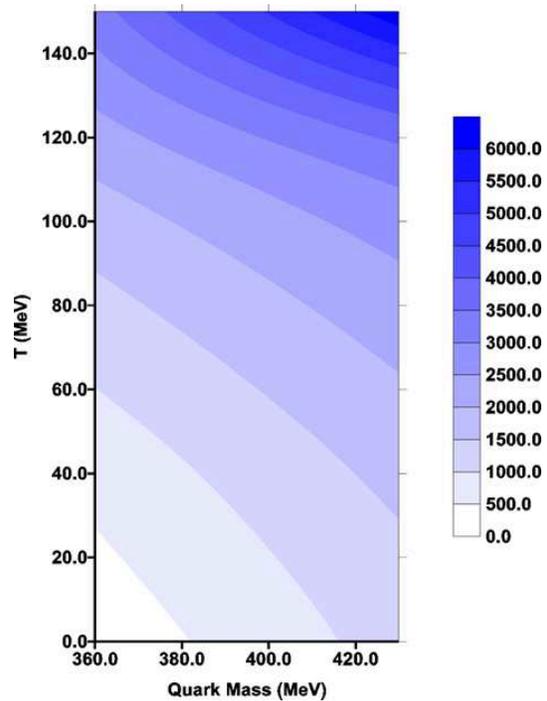}
\vskip-3mm\caption{ Dependence of the higher-order effective
potential on the quark mass and a finite temperature }\vskip2mm
\end{figure}%
\begin{figure}%
\includegraphics[width=7cm]{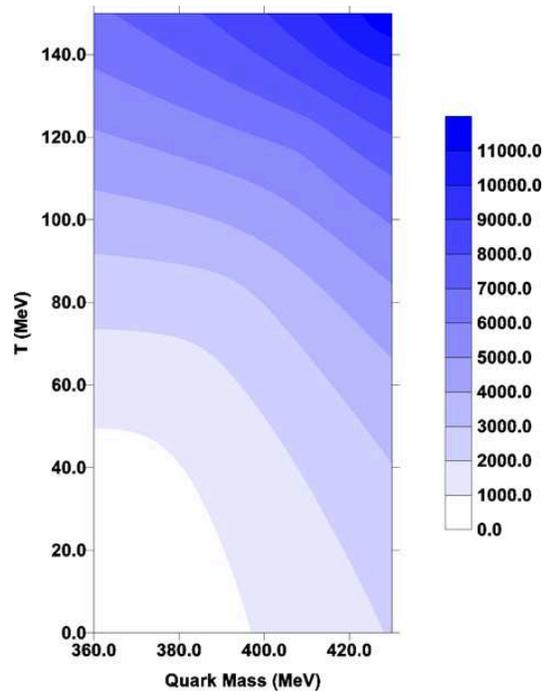}
\vskip-3mm\caption{ Dependence of the normal effective potential on
the quark mass and the finite temperature }
\end{figure}%
Akoi {\it et al.} [26] performed lattice calculations and found that
the phase transition is a crossover in the absence of a chemical
potential. We find that the phase transition is a crossover when
higher-order mesonic interactions are included. In Ref. [23], the
authors investigated the Goldstone theorem below the critical
temperature in the linear sigma model in the chiral limit. They
found that the Goldstone theorem is satisfied, which is in agreement
with the present results.

\section{Summary and Conclusion}

The chiral phase transition, the sigma and pion masses, and the
effective mesonic potential have been examined in the framework of the
extended linear sigma model, in which higher-order mesonic
interactions are taken into account. We calculated the sigma and
pion masses by minimizing the potential. This technique has been
used in previous studies such as Ref. [22]. We find that the
critical temperature is reduced to lower values of the temperature
when the higher-order mesonic interactions are included in the
linear sigma model, leading a good agreement with lattice QCD [26].
We find that the Goldstone theorem is satisfied in the chiral limit
at low temperatures. A comparison with the other calculations and
lattice QCD calculations is presented. The effective sigma and pion
masses are investigated in the chiral limit and in the presence of
an explicit symmetry-breaking term. Moreover, we find that the second order phase transition becomes
a crossover in the chiral limit, which agrees with lattice QCD
[26]. Hence, we conclude that the higher-order mesonic interactions
play an important role in a change of the meson properties at
finite temperatures.

\vspace*{-5mm}
\rezume{%
М.\,Абу-Шаді, Х.М.\,Мансур} {ВПЛИВ МЕЗОННИХ ВЗАЄМОДІЙ\\ ВИСОКИХ
ПОРЯДКІВ НА КІРАЛЬНИЙ ФАЗОВИЙ\\ ПЕРЕХІД І КРИТИЧНУ ТЕМПЕРАТУРУ} {У
даній роботі мезонні взаємодії високих порядків враховано в лінійній
сигма-моделі при кінцевій температурі. При обчисленні мас
сигма-частинки і піона ефективний потенціал мінімізується. Рівняння
поля вирішені в наближенні середнього поля розширеним методом
ітерацій при кінцевій температурі. Порядок кірального фазового
переходу, ефективні маси сигма-частинки і піона і ефективний
мезонний потенціал досліджені як функції температури. Знайдено, що
кіральний фазовий перехід задовольняє теорему Голдстоуна нижче
критичної точки, коли умову мінімізації виконано в кіральній
границі. Величина критичної температури менше порівняно з вихідною
моделлю у згоді з результатами ґраткової КХД. Проведено порівняння
модифікованої моделі з моделями інших робіт.}

\end{document}